# Fetal Head and Abdomen Measurement Using Convolutional Neural Network, Hough Transform, and Difference of Gaussian Revolved along Elliptical Path (Dogell) Algorithm

Kezia Irene, Aditya Yudha P., Harlan Haidi, Nurul Faza, Winston Chandra
Faculty of Computer Science Universitas Indonesia Kampus Baru UI Depok – 16424 Jawa Barat, Indonesia
Email: kezia.irene@.ui.ac.id

*Abstract*—**The number of fetal-neonatal death in Indonesia is still high compared to developed countries. This is caused by the absence of maternal monitoring during pregnancy. This paper, we present an automated measurement for fetal head circumference (HC) and abdominal circumference (AC) from ultrasonography (USG) image. This automated measurement is beneficial to detect early fetal abnormalities during the pregnancy period. We used the convolutional neural network (CNN) method, to preprocess the USG data. After that, we approximate the head and abdominal circumference using the Hough transform algorithm and the difference of Gaussian Revolved along Elliptical Path (Dogell) Algorithm. We used the data set from national hospitals in Indonesia and for the accuracy measurement, we compared our results to the annotated images measured by professional obstetricians. The result shows that by using CNN, we reduced errors caused by a noisy image. We found that the Dogell algorithm performs better than the Hough transform algorithm in both time and accuracy. This is the first HC and AC approximation that used the CNN method to preprocess the data.**

## I. INTRODUCTION

The Indonesian Central Bureau of Statistics reported that in 2017, Indonesia has reached the lowest point of fetal-neonatal death, which is 1.5% [1]. This number has gone much lower over a ten-year period. Where in 2007, the number of fetal-neonatal death was 3.4% [2]. However, the rate of fetal-neonatal death in Indonesia is still bad compared to other countries. According to Unicef [3], in Japan, the rate of fetal-neonatal death in 2017 is only 0.1%, while in the US, the fetal-neonatal death rate in 2017 is 0.4%.

According to the Ministry of Women's Empowerment and Child Protection [4], 62.56% of childbirth in Indonesia is assisted by a midwife, while only 30% assisted by an obstetrician. Those people who only went to a midwife never got a modern treatment to detect their fetus abnormalities. The big number of people going to a midwife instead of an obstetrician is because of the lack of obstetricians in rural areas [4]. Most of the obstetricians are concentrated in big cities, towns, and capital.

Since 45% of the people in Indonesia live in rural areas [6], the lack of obstetricians has caused the absence of monitoring during the pregnancy process. The absence of monitoring during pregnancy has caused many predicaments, especially during childbirth [7]. These predicaments include maternal deaths, maternal morbidity, and fetal deaths [7].

In order to solve this problem, we use a telehealth monitoring system [8]. This system will allow people from different locations to exchange their health data [8]. We aimed to give access to people who live in remote locations to be able to send their health data to the professionals without having to travel to get their body checked up. This system has also automated some of its features. One of them is fetal organ detection from ultrasonography (USG) images.

Fetal organ detection from USG images includes the detection of head circumference, biparietal diameter, abdominal circumference, femur length, and humerus length [8]. This detection works in all three stages of pregnancy, the first, second, and third trimester. However, due to the nescience knowledge of the best algorithm, currently, the best accuracy was 86.42% ± 11.69 [9]. Therefore, professional assistance after the approximation of a USG image is still needed.

This paper focuses on measuring the head circumference (HC) and abdomen circumference (AC) from a USG image. Many algorithms have been proposed for computing the circumference and the diameter of ellipse images. Some researches used unsupervised methods, such as flower pollination algorithm and cuckoo search [9], while some others used a combination of supervised and unsupervised methods like constrained probabilistic boosting tree [10][11].

The state of the art in this paper is that we use a convolutional neural network algorithm, YOLO (You Only Look Once) to pre-process the data. After the preprocessing step, we use both the Hough Transform algorithm and The Difference of Gaussian Revolved along Elliptical Path (Dogell). The preprocessing step on processing the data allows the reduction of errors caused by wrong ellipse detection. Moreover, this paper will also predict if there is an abnormality based on a comparison with the international fetal growth standard [12].

## II. LITERATURE REVIEW

There have been many studies conducted that aim the same goals. We have reviewed several studies that used both the same and different methods. We only focus on the paper that has the same goal with us, which is to approximate the head and abdominal circumference and also the biparietal diameter.

A study by Jatmiko, W., et al [13] used AdaBoost.MH classifier to preprocess the image, combined with the Hough transform algorithm to approximate the ellipse circumference. AdaBoost.MH is a supervised learning algorithm that requires a lot of training data. After using the AdaBoost.MH classifier, they used the Haar-like feature extraction to superimpose several rectangular regions over the image, and calculate the differences between each region. However, the steps needed for this approximation is very complex that makes the space complexity quite high.

Van den Heuvel, et al [15] also used a similar method. They first used the random forest classifier algorithm to compute the Haar-like features. After that, they extracted the Haar-like features using Hough transform, dynamic programming, and ellipse fit. They used a large test set which contained data of all trimesters of the pregnancy. Their overall accuracy was 0.8 ± 2.6 for the first trimester, 0.0 ± 4.6 for the second trimester, and 1.9 ± 11.0 for the third trimester.

Carneiro, et al [10] used constrained probabilistic boosting tree algorithm to approximate the fetal anatomical structures USG images. After segmenting their images using the constrained probabilistic boosting tree, they used mathematical equations to approximate the biparietal diameter, head circumference, and abdominal circumference. However, due to the novelty of their method, the overall accuracy still needs to be improved.

Ma'sum, M. A., et al [16] approximated the fetal head circumference and biparietal diameter using the difference of Gaussians revolved along elliptical path (Dogell) algorithm along with particle swarm optimization (PSO). They found that using the Dogell algorithm combined with PSO algorithm results in the better accuracy of the approximation.

The more modern approach has been studied by Sofka, M, et al [17]. They measured the fetal head ultrasound using sequential estimation and integrated detection network (IDN). The study result is an average difference of 2 mm between the ground truth and the automatic measurement with the running time of 6.9 s (GPU) or 14.7 s (CPU).

## III. RESEARCH METHODOLOGY

### *A. Data Labeling*

Before we process the datasets, we need to get some features and labels from them. We used the image as to features. In order to label our data, we used an image labeling software called LabelImg. We saved the labeled data in XML file. The XML file will be used as input in darkflow. Darkflow is one of the tools that implement YOLO algorithm. We used darkflow to train the model by executing flow command followed by our model, load (weights), train, annotation, and dataset options.

### *B. You Only Look Once (YOLO)*

You Only Look Once the algorithm was developed by Redmon et al [14]. YOLO algorithm is an object detection that implements a Convolutional Neural Network (CNN). YOLO detects the separate components of object detection into a single neural network [14]. Our model detects 2 objects (Head Circumference and Abdomen Circumference).

YOLO system divides the input image into an S x S grid. If the center of an object falls into a grid cell, that grid cell is responsible for detecting that object. Each grid cell predicts B bounding boxes and confidence scores for those boxes. Confidence define as Pr(Object)*IOU [14]. Confidence scores should be zero if no object exists in that cell. Each grid cell also predicts C conditional probabilities, $Pr(Class_i|Object)$. These probabilities are conditioned on the grid cell containing an object. The system only predicts one set of class probabilities per grid cell.

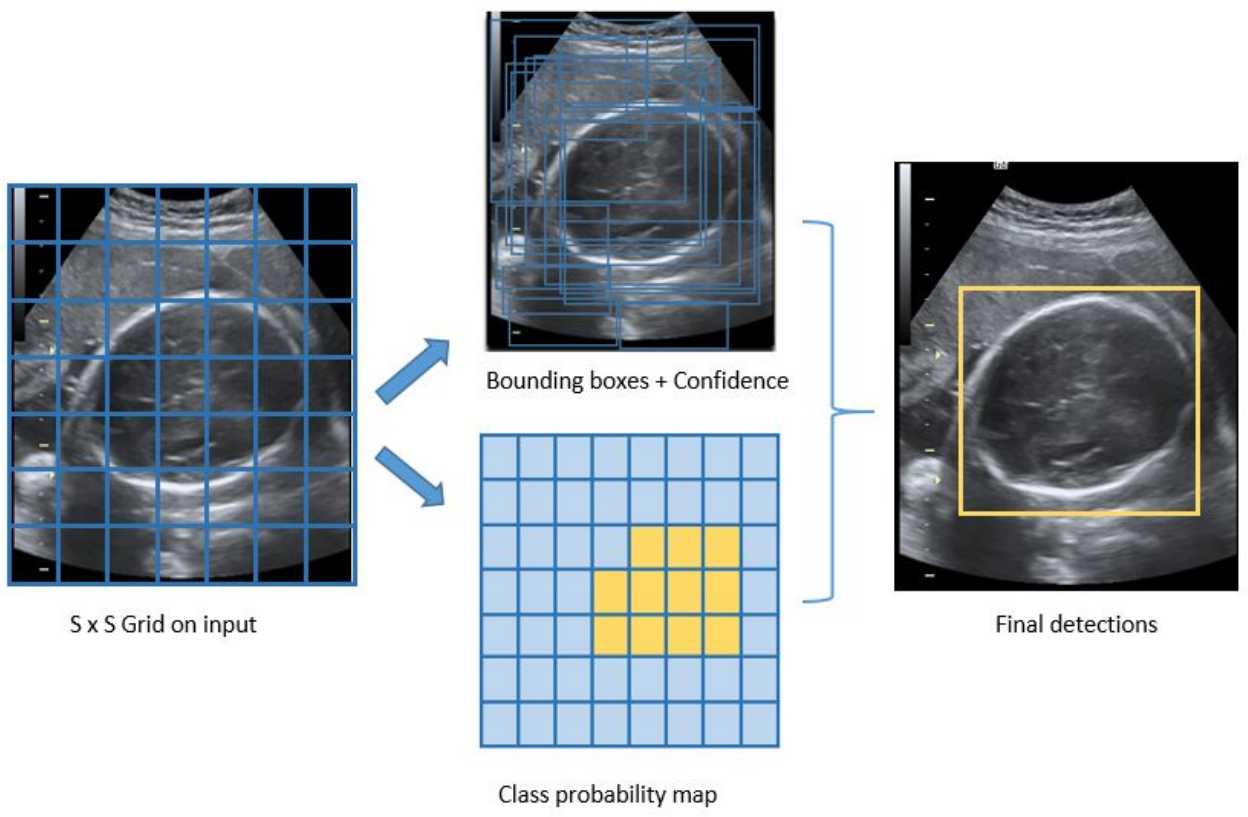

Fig. 1. The Model

System models detection as a regression problem. It divides the input into an S x S grid and for each grid cell predicts B bounding boxes, confidence for those boxes, and C class probabilities.

YOLO implements this model as a convolutional neural network. The architecture of the YOLO network has 24 convolutional layers followed by 2 fully connected layers.

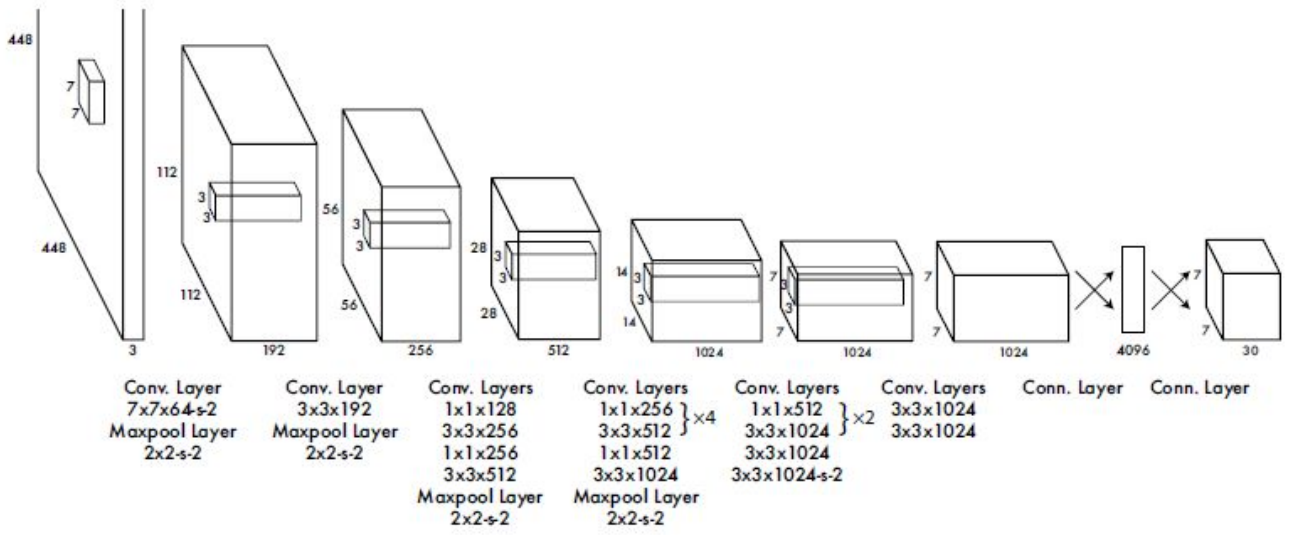

Fig. 2. YOLO Architecture, proposed by Redmon et al [14]

The reason why we choose YOLO as object detection, because we want to use the fastest CNN algorithm and because every image contain exact 2 circles (for head and abdomen), so if we check the image once, it is already enough to get the marker boundary box. After we get the result from YOLO, we crop the image and use the segmented image as input in Hough Transform or DoGell.

### *C. Hough Transform*

Hough Transform is a robust tool to extract features from images and describe them parametrically. The idea of Hough Transform for shape detection is by transforming the spatial domain into "parameter space" where the characteristics of the particular shape we want to find can be voted to determine its actual position.

We used the hough transform method from the scikit-image library. First, we converted the RGB image of segmented USG to grayscale and detected the edges. Then, we performed hough transform to the image using hough_elips method and got the results. We chose the best ellipse result which located at the end of the sorted result array. After that, we draw the ellipse on the image.

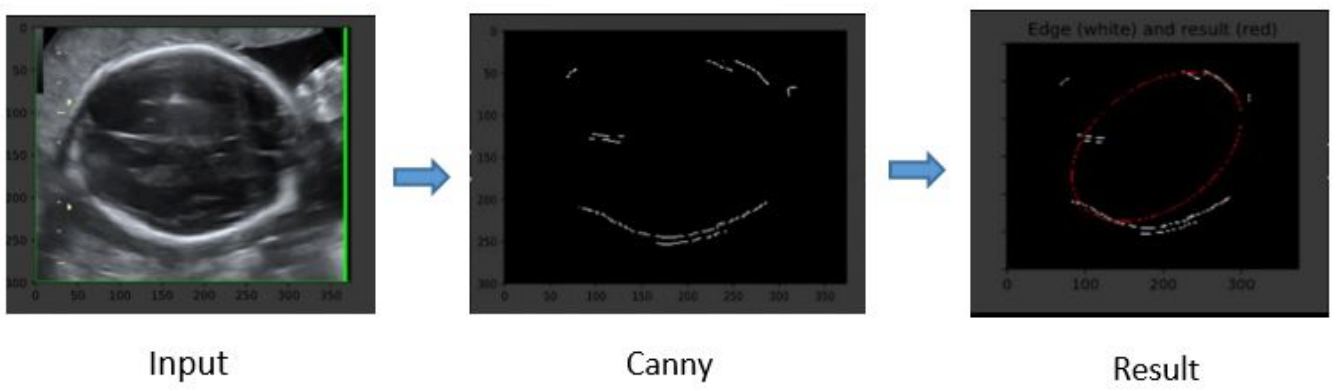

Fig. 3. Hough Transform Process

### *D. The difference of Gaussian Revolved Along Elliptical Path (DoGell)*

The difference of Gaussian revolved along an elliptical path (DoGell) is a model proposed by Foi et al [18]. Dogell is a method for segmenting fetal head circumference. Dogell assumes that the shape of the object is an ellipse. Detected ellipse size can also be calculated by dogell method.

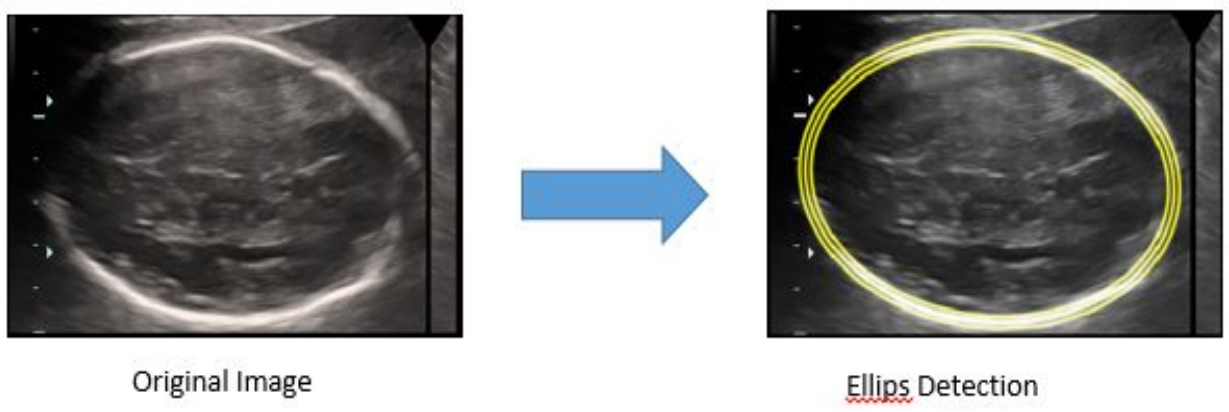

Fig. 4. Ellipse Detection with DoGell

From the ellipse model, the gaussian ellipse surface computed using equation (c1 and c2 = coordinates of ellipse center, r1 and r2 = mayor and minor axis, theta = ellipse orientation, and t = ellipse thickness):

$$g(x_1, x_2, \boldsymbol{a}, s) = \frac{f_s(d(x_1, x_2, \boldsymbol{a})) - f_{3s}(d(x_1, x_2, \boldsymbol{a}))}{r_0(x_1, x_2, \boldsymbol{a})}$$

Fig. 5. Gaussian Ellipse Surface Computation [16]

then, the score of the ellipse surface for the given ultrasound image is computed using the equation:

$$C(z, \boldsymbol{a}, s) = \iint z(x_1, x_2) g(x_1, x_2, \boldsymbol{a}, s)\, dx_1 dx_2 + \lambda \left( \max\left(0, \frac{\max(r_1, r_2)}{\min(r_1, r_2)} - CI\right)\right)^2$$

Fig. 6. Score Ellipse Surface Computation [16]

after both of the above computations done, the result of ellipse detection can be plotted and visible to be seen.

## IV. RESULTS AND ANALYSIS

### *A. Dataset*

The dataset used for this paper is the first and second-trimester ultrasound images. This dataset consists of 326 images with size 800 x 600 pixels. Each image contains both the fetal head and abdomen as shown in Figure 1. There are also specks of noise in images, which can lead the experiment to failed approaches if they have high intensity. On the left side of the image, there is a gray-scale level bar to show shades of gray available in the image. This dataset has an annotated version that can be used to evaluate the approximation based on the actual results from the medical expert.

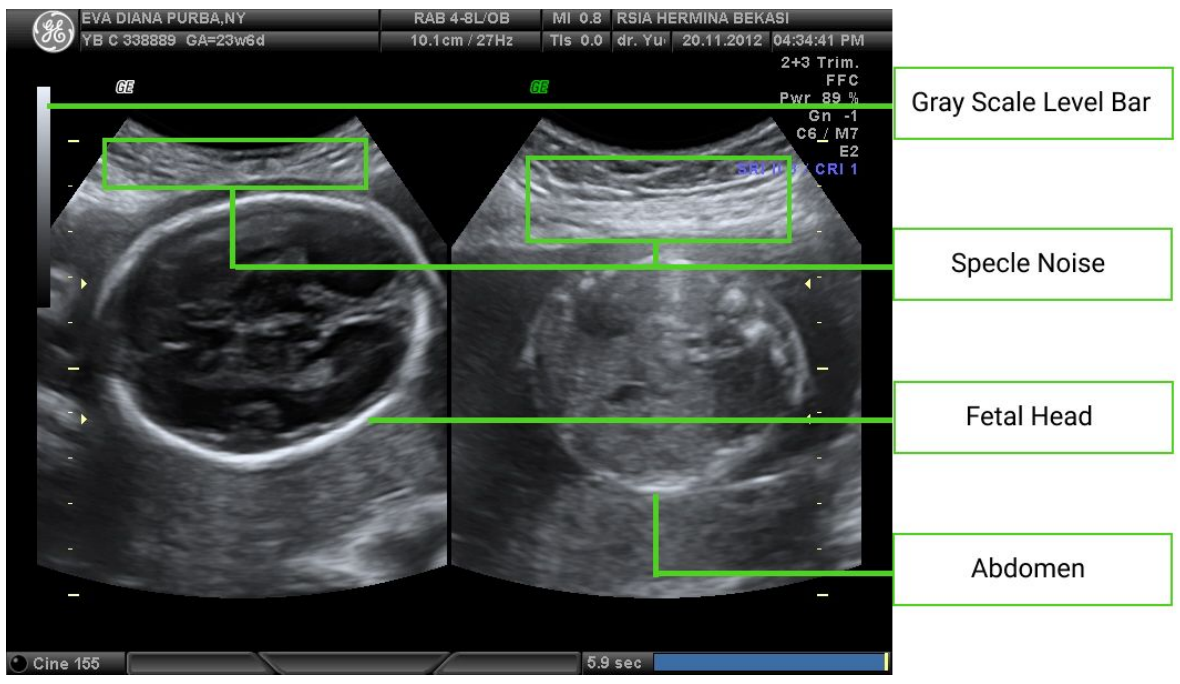

Fig 7. USG Dataset

### *B. Perimeter Approximation*

Fetal head and abdomen ellipses are detected using YOLO. Detected objects are marked with a green boundary box. After that, we segmented the image based on the boundary box into two images, abdomen, and fetal head. This segmentation is done so that the ellipse used for approximating the perimeter will be more specific. This also prevents the detection of circles outside the abdomen and fetal head that will affect the results and reduce the accuracy.

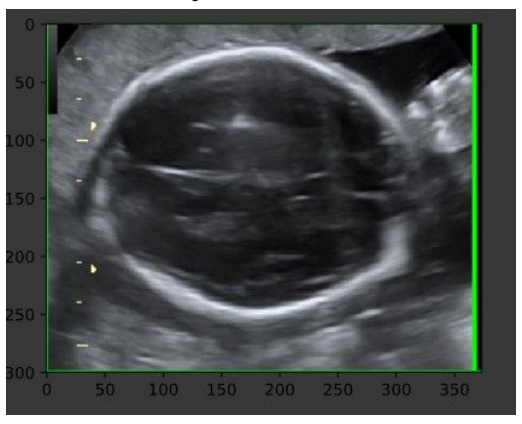

Fig.8. Segmented Fetal Head Circumference

We used the Hough Transform and DoGell algorithm to detect the ellipses of the fetal head and abdomen in 38 ultrasound images. Hough Transform detected ellipse's canny edges and then formed the ellipse based on its edges (Figure 3). The perimeter of the ellipse in the Hough Transform is calculated using its major and minor axis information.

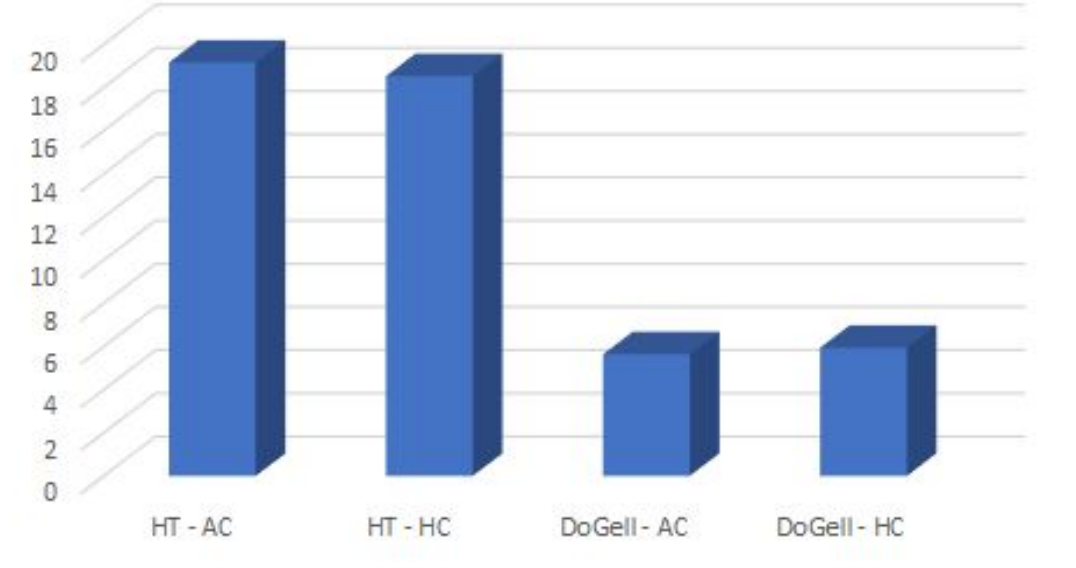

Fig.9. Mean Absolute Difference of Head and Abdomen Circumference using Hough Transform and DoGell Algorithm

TABLE I. MEAN ABSOLUTE DIFFERENCE RESULT

| **Method** | **Mean Absolute Difference** | |
|---|---|---|
| | ***Head Circumference*** | ***Abdomen Circumference*** |
| Hough Transform | 18.52594877 | 19.14858314 |
| DoGell | 5.937809729 | 5.638809099 |

Results in both methods are evaluated by calculating their mean absolute difference (MAD) with the actual perimeter. As shown in Table 1, the MAD result for the Hough Transform is very large compared to the expert's data. While DoGell algorithm has more precise results in approximating head and abdomen circumference.

We found approximation results in the Hough Transform method precise for one image. Turns out, it happens because the tuning parameters we did in Hough Transform referred to this one image. Therefore, it can be concluded that the large MAD in the Hough Transform method is due to tuning parameters that cannot be used generally for all testing data.

## V. CONCLUSION

In this paper, we have approximated the head circumference (HC) and abdominal circumference (AC) from fetal ultrasound images. We used a convolutional neural network (CNN) to preprocess the data and then approximate the HC and AC using the Hough transform algorithm and the difference of Gaussians revolved along elliptical paths (Dogell) algorithm. After the experiment, we found that CNN can reduce the error caused by noisy background and multiple ellipses in the same picture. We also found that after preprocessed by the CNN algorithm, the accuracy of the Dogell algorithm is higher than the Hough transform algorithm. Dogell algorithm is also faster compared to the Hough transform algorithm by 3.1646 s (CPU).

## ACKNOWLEDGMENT

This work is supported by the Faculty of Computer Science, Universitas Indonesia, supervised by M. Anwar Ma'sum and Ari Wibisono.